\RequirePackage{fix-cm}
\documentclass[twocolumn]{svjour3}          
\smartqed  
\usepackage{graphicx}
\usepackage{float}
\usepackage{pdfpages}
\begin{document}
\title{Hybrid time series from PIV for characterization of turbulent flow fields 
}
\subtitle{ASTRA - Approach using Spatially and Temporally Resolved Advection}


\author{Tom T. B. Wester          \and
        Andr\'{e} Fuchs          \and
        Joachim Peinke          \and
        Gerd G\"ulker
}


\institute{T. T. B. Wester \at
              K\"upkersweg 70, 26129 Oldenburg \\
              Tel.: +49-0441-798 5023\\
              Fax: +49-0441-798 5099\\
              \email{tom.wester@uni-oldenburg.de}           
}

\date{Submitted to ArXiv on 23.05.2021}

\maketitle

\begin{abstract}
Particle Image Velocimetry (PIV) has become increasingly popular to study structures in turbulent flows. PIV allows direct extraction and investigation of spatial structures in the given flow field. Increasing temporal resolution of PIV systems allows a more accurate capture of the flow evolution. Despite the very good spatial resolution of PIV, current systems can only match the multiple $kHz$ sampling rates of hot-wire or Laser Doppler Anemometer (LDA) measurements for a very short period in temporal analyses of flow.\\
In this study, an advection-based approach is presented which uses Taylor's hypothesis of "frozen turbulence" for small scale turbulent patterns. Compared to the underlying raw data a major increase of the temporal resolution for extracted time series is shown. The quality of the presented approach is shown for two-point analyses, which would not be possible with presently known methods. To demonstrate this, different turbulent flow cases behind a fractal grid are studied. For the validation of the results corresponding hot-wire measurements at various positions along the centerline were used.\\
\end{abstract}

\section{Introduction}
\label{intro}
In recent years, a remarkable development in the field of Particle Image Velocimetry (PIV) can be observed \cite{raffel2018particle}. The most important evolutions are the faster hardware components, in particular lasers and cameras \cite{Hain2007}. Current systems allow continuous frame rates of 1kHz up to 10kHz. Higher frame rates are mainly inaccessible due to the needed illumination power. Only in exceptional cases it is possible to achieve significantly higher sampling rates. An example is the so-called postage stamp PIV with frame rates as high as 400kHz \cite{beresh2018postage}. However, this can only be done for a handful of images and very small field of view (FOV). Consequently, those PIV measurements can only give a brief glimpse of the flow structures using pulse-burst lasers \cite{Beresh2015}. However, there are several experimental situations where high frequencies above 40kHz are needed most prominently for transonic flows \cite{Pope2001}. But also, very high Reynolds numbers or the choice of different fluids may lead to very small non-viscous turbulent flow structures and correspondingly high frequencies.\\
PIV captures the entire flow field with at least two velocity components. This is of clear advantage in many applications compared to point measurement methods such as hot-wire or Laser Doppler Anemometer (LDA). Using time resolved PIV each interrogation window 
(IW)\footnote{The raw PIV images are divided into smaller areas called interrogation windows. In further analysis, the interrogation windows of successive images are correlated to determine the particle displacement and thus the velocity. The size of the interrogation window determines the spatial accuracy and the smoothing of the resulting flow field \cite{raffel2018particle}.} 
can be taken as a single point measurement of investigated flow field. Although the resulting spatial resolution is usually of the same order of magnitude, it is nevertheless difficult to match the temporal resolution of hot-wire or LDA measurements.\\
To artificially increase the temporal resolution of PIV measurements Scarano and Moore proposed a new approach almost a decade ago \cite{Scarano2012}. Instead of using only the temporal or only the spatial information embedded in PIV measurements, they proposed an approach of ''Pouring Space in Time'' (in the following also referred to as \textit{PST}) based on Taylor's hypothesis of ''frozen turbulence''. With \textit{PST} velocity fields of two successive PIV images are interpolated to increase the actual frame rate of a given PIV measurement. This approach is a further development of \cite{Tennekes1975}, \cite{foucaut2004piv} and \cite{DelAlamo2009}, who showed the transformation from spatially measured structures into the temporal domain by Taylor's hypothesis. \cite{DelAlamo2009} also showed the justification of this approach for small spatial distances.\\
In their approach Scarano and Moore estimate the inter frame movement of the structures by the linear part of the Taylor expansion of the complex fluid movement. Thus, for each IW the flow propagation for a time step $\Delta$t is determined. Using this approach, a forward evolution of the flow field is calculated between two successive images for image 1 and a backward evolution for image 2. The two resulting flow fields are averaged using a weighting factor. The resulting intermediate flow fields lead to the increased frame rate. The upper limit of temporal resolution is given by the spatial resolution, since the fluid motion must be larger than an interrogation window. A similar approach was also introduced in \cite{Kat2012}. Scarano and Moore's idea was later also generalized to volumes \cite{Schneiders2014}. The demand of such an increase in temporal resolution has shown to be very important for the extraction of pressure gradients from PIV data where the acceleration must be known precisely \cite{Liu2006,Haigermoser2009,Charonko2010,Violato2011} or as an input for simulations \cite{Suzuki2012,Gronskis2013,Lemke2013,Suzuki2014,Vlasenko2015}. \\
In the present study we will show a comparable approach as proposed by Scarano and Moore, but without the need of interpolation in between frames. The \textbf{A}pproach using \textbf{S}patially and \textbf{T}emporally \textbf{R}esolved \textbf{A}dvection (or short \textit{ASTRA}) also uses Taylor's hypothesis of frozen turbulence to connect the temporal and spatial information contained in the PIV data. While Scarano and Moore estimate the propagation of the flow, we assume all needed information of the flow field are already contained in the neighboring IWs. Thus, by simply extracting the given data, the sampling rate can be significantly increased, depending on flow velocity and spatial resolution. While PST calculates entire velocity fields, our approach is currently limited to individual local time series. However, a direct comparison shows \textit{ASTRA} to preserves the statistics of small-scale events, while this is lost in other methods. In the field of turbulence research, it is precisely these small structures which are of great importance.\\
To verify the new approach, the complex turbulent flow field behind a fractal grid is studied. In Sec. \ref{sec:Setup} the used setup is described. In Sec. \ref{sec:Methods} the most important approaches using PIV data are introduced. Following in Sec. \ref{sec:HybApp} the method of \textit{ASTRA} is presented. Here also the influence of the PIV evaluation parameter on the results is investigated. Sec. \ref{sec:Results} will present the results generated by the method and compares them against hot-wire measurements, as well as the \textit{PST} approach of Scarano and Moore. The paper ends with a conclusion in Sec. \ref{sec:Conclusion}.

\section{Experimental Setup}
\label{sec:Setup}
\begin{figure*}[htb]
	\begin{centering}
		\includegraphics[width=0.7\textwidth]{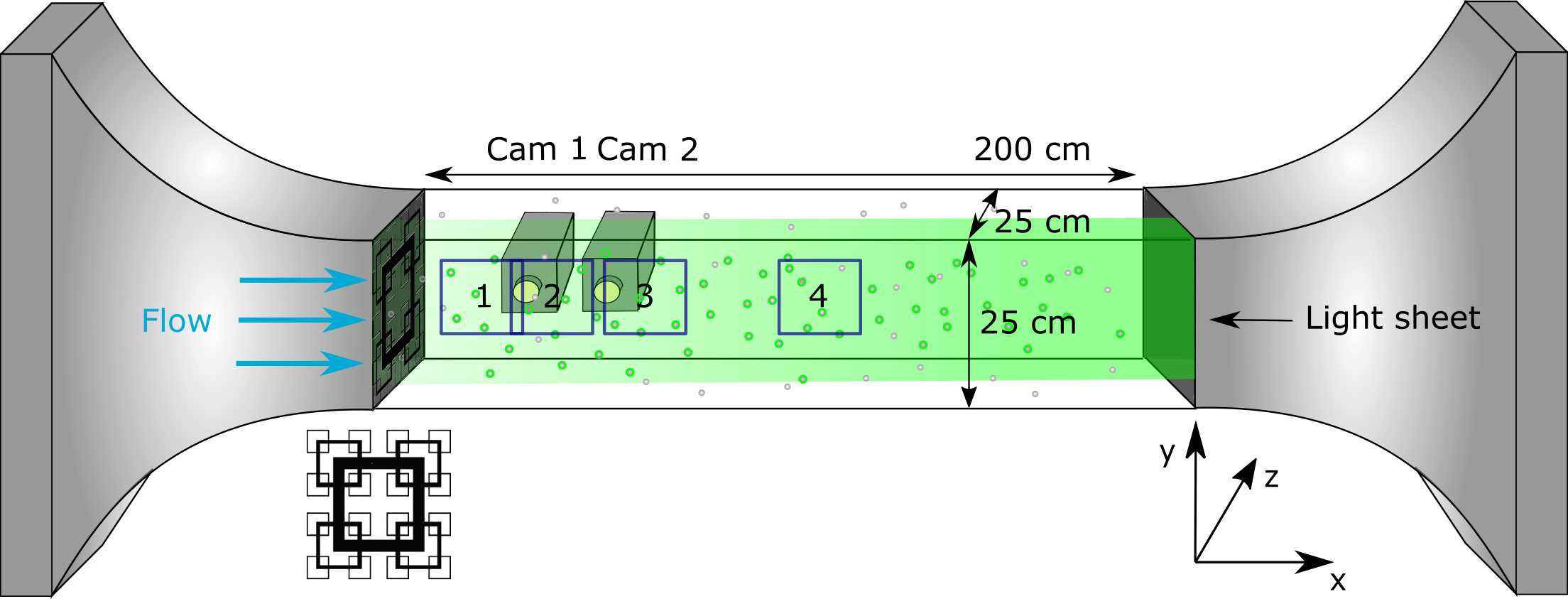}
		\caption{Schematic sketch of the used setup including the PIV light sheet, PIV cameras, four measurement positions and fractal grid. The measurement positions are chosen to investigate both the complex region near the grid and the fully developed turbulence in the far field.}
		\label{fig:Setup}       
	\end{centering}
\end{figure*}

\begin{figure*}[h]
\begin{centering}
	\includegraphics[width=0.75\textwidth]{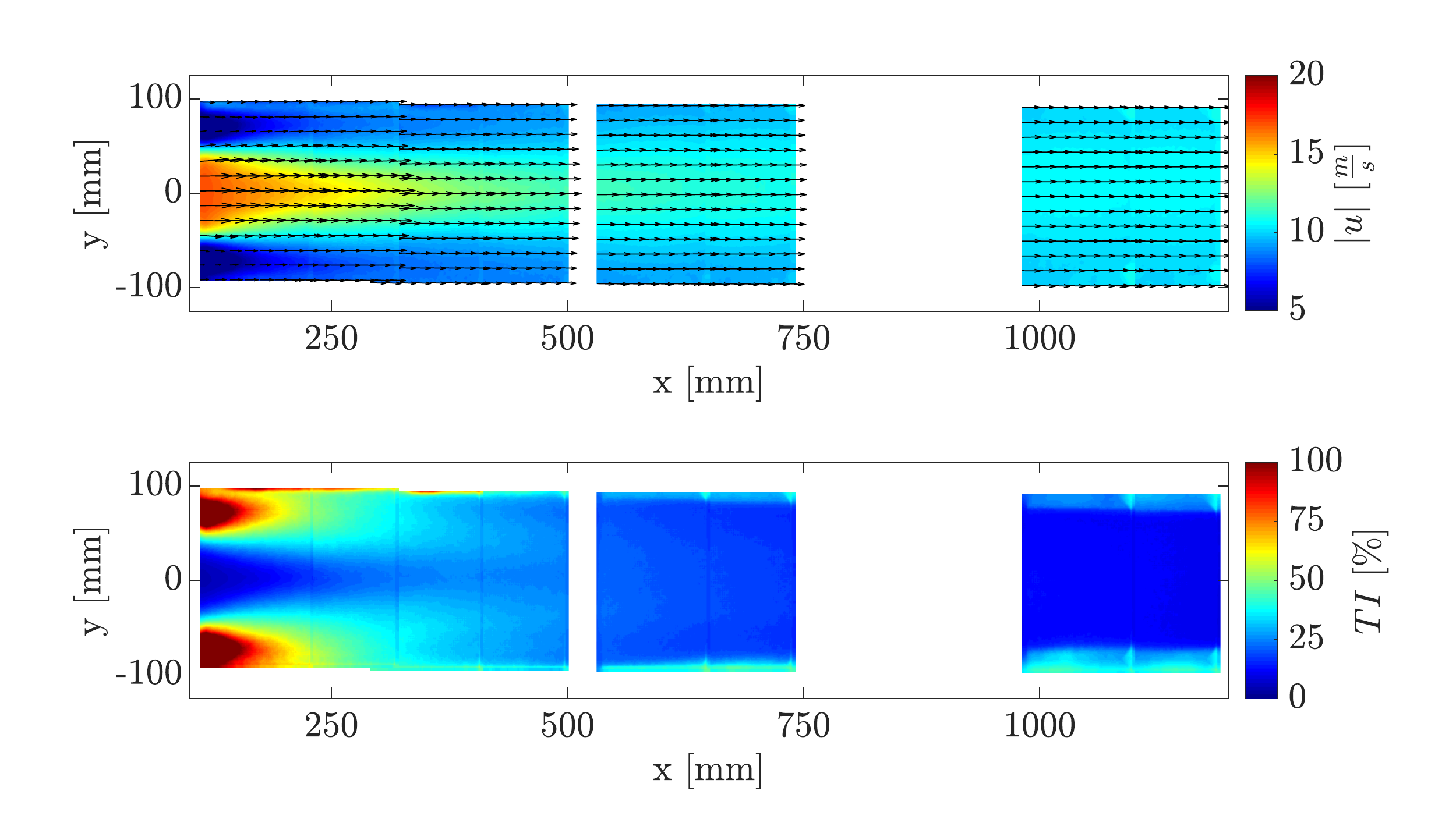}
	\caption{Top: Average flow field behind the fractal grid composed of the four PIV positions. The absolute velocity is represented color coded in the background. The arrows represent the local flow direction. Bottom: Turbulence intensity of the flow behind the fractal grid.}
	\label{fig:PIVField}       
\end{centering}
\end{figure*}

To validate the \textit{ASTRA}, a turbulent flow behind a fractal grid is used. The experiments are conducted in a G\"{o}ttingen type return wind tunnel at the University of Oldenburg. The wind tunnel has a $0.25$m $\times$ $0.25$m $\times$ $2.00$m (height $\times$ width $\times$ length) test section made of acrylic glass to allow optical access from nozzle to funnel. The wind tunnel has a glass window in the funnel to allow a light sheet to be coupled into the measuring section from downstream and thus illuminate the entire cross-section. A schematic sketch of the setup is shown in Fig. \ref{fig:Setup}, where the flow comes from the left-hand side. During the experiments, the free stream velocity is set to $u_\infty = 10 \frac{m}{s}$. \\
A fractal grid is used to generate a versatile flow. This flow shows a very coherent structure due to the meandering jet, as well as a very inhomogeneous and highly dynamic flow field near the grid. Further downstream, a very homogeneous turbulence results, without any coherent structures remaining. The used fractal grid has $N=3$ iterations and is based on a square shape \cite{seoud2007dissipation,valente2011decay,mazellier2010turbulence}. The sizes of the squares of subsequent iterations are characterized by the ratio $R_L = \frac{L_j}{L_{j-1}}=0.52$ and for the bar thickness by the ratio $R_t = \frac{t_j}{t_{j-1}}=0.36$ with the scales of the first iteration $L_0 = 138.4 $mm and $t_0 =20.1$mm. With given parameters the grid has a blockage of $38.2 \%$. The grid has an overall size of $0.25$m $\times$ $0.25$m and therefore fits exactly into the outlet of the nozzle where it is placed (see Fig. \ref{fig:Setup}).\\
To get well-resolved time series for comparison, hot-wire measurements are performed along the centerline. Positions with a stream wise distance of $50$mm $\leq x \leq 1760$mm behind the grid and a step size of $20$mm are chosen. The used hot-wire is a single wire probe of type $55P01$ with a length of $l_{HW} = (2.0 \pm 0.1)$mm and a diameter of $d_{HW} = 5\mu $m from Dantec Dynamics together with a StreamLine frame with a $90C10$ CTA module. For calibration of the hot-wire a \textit{Dantec Dynamics Hot-Wire Calibration Unit} is used. The software used for capturing the data is the StreamWare version $3.50.0.9$. The measurements are taken with a sampling frequency of $f_{HW} =60$kHz with a \textit{NI PXI 1042} AD-converter for a duration of 60s per point. To satisfy the Nyquist theorem the data are low pass filtered at a frequency of $30 $kHz. \\
A high-speed PIV system is used in a two-component (2C) two-dimensional (2D) mono setup to characterize the flow in a spatial manner. To illuminate the measurement area, a \textit{Litron 303HE} Nd:YLF Laser ($\lambda = 527 $nm) is used. A light sheet is formed and positioned mid-plane of the test section. For flow seeding DEHS droplets of a size of $\approx$1 $\mu$m are used. The sampling frequency of the PIV system is set to $F_{s,PIV} = 693 Hz$ corresponding to the full resolution frame rate for double images of the used \textit{Phantom Miro 320S} cameras. The captured images have a resolution of 1920 $\times$ 1080 px. During the experiment two cameras are used side by side each fitted with \textit{Nikon mikro NIKKOR 55}mm objectives with an aperture setting of f/8. 
Cameras and laser are synchronized using the \textit{LaVision programmable timing unit (PTU)}. The PIV processing is done using the commercial \textit{LaVision DaVis 8.4} software. The velocity vector fields are calculated using the recursive cross-correlation engine. Different IW sizes from 12 $\times$ 12 px up to 64 $\times$ 64 px and IW overlaps from 0\% up to 75\% are used to examine the influence of the PIV evaluation parameter on the approach. Since these parameters also determine the spatial resolution $R$ of the calculated vector fields, $R$ is between $0.29 \times 0.29$ mm$^2$ and $1.6 \times 1.6$ mm$^2$.\\
Given setup allows the measurement of a field of view (FOV) of approximately 209mm $\times$ 189mm (x $\times$ y). A total of four positions are investigated to characterize the flow. Three of them are in the dynamic range directly behind the grid. The fourth position is downstream in the far field to capture also the fully developed flow (see Tab. \ref{tab1} and Fig. \ref{fig:PIVField}). For each of the shown measurement positions, six data sets of 1822 image pairs are recorded. The number of image pairs is given by the camera memory. Fig. \ref{fig:PIVField} shows the resulting mean flow field and turbulence intensity calculated from the measurements. Due to the homogeneous turbulence the fourth position is used to determine the optimal evaluation parameters of the method (see Sec. \ref{sec:PIV_Param}).

\begin{table}[htbp]
	\centering
	\caption{Field of views (FOV) of PIV measurements}
	\begin{tabular}{|c|c|c|c|c|}
		\hline 
		& $x_{min}$ /& $x_{max}$ / & $y_{min}$ /& $y_{max}$ /\\
		& mm & mm & mm & mm \\ 
		\hline 
		Position 1 & 111.2 & 321.1 & -91.3 & 97.7 \\ 
		\hline 
		Position 2 & 291.2 & 501.1 & -94.3 & 94.7 \\ 
		\hline 
		Position 3 & 531.2 & 741.1 & -95.3 & 93.7 \\ 
		\hline 
		Position 4 & 981.2 & 1191.0 & -97.3 & 91.7 \\
		\hline 
	\end{tabular}
	\label{tab1} 
\end{table}%

\section{Structures from PIV Fields}
\label{sec:Methods}
Since PIV was first introduced, the velocity fields obtained have been utilized to extract structures and temporal information. The most used approaches will therefore be briefly introduced here.\\
If the temporal resolution of a PIV measurement is rather poor, only the spatial information is extracted from the independent measured velocity fields. To obtain the most information from the measurement nevertheless, the entire FOV of the PIV measurement is usually included in further analysis. The high spatial resolutions of the velocity fields provide a very detailed insight into the structures \cite{foucaut2004piv,Tennekes1975,stanislas1998experimental}. This approach provides spatially high resolved data but is limited by the spatial range of the vector field resolution. Structures larger than the FOV under investigation cannot be resolved. In addition, in case of inhomogeneous flows, a mixing of different statistics occurs, because of the large area under consideration. Due to these disadvantages and todays temporally better resolved PIV systems, this method is rarely used anymore. For this reason, this approach will not be discussed further in this paper.\\
Nowadays, the temporal resolution of PIV systems enables a direct extraction of well resolved time series. Here, each interrogation window is interpreted as a single probe in the flow. Therefore, temporally resolved data can be extracted locally at any point of the measurement at the sampling rate of the PIV system itself. This prevents spatial averaging of statistical features and allows structures larger than the FOV to be captured. However, the smaller structures are often lost due to the temporal resolution. This approach will serve as the baseline case in the following and is referred to as \textit{PIV}.\\
Since, despite increasing temporal resolution, there is still a large gap between PIV and hot-wire or LDA, an advection-based approach has been presented in the past by Scarano and Moore. It is called "Pouring Space into Time"\cite{Scarano2012}. By interpolation of successive vector fields, any new sampling rate can be selected here, with certain limitations. Due to the interpolation in between frames whole PIV fields are calculated. In the following, this approach will be referred to as \textit{PST}. Since this is still the most common and commercially available method to resample PIV data in time, it will be compared with our new \textbf{A}pproach using \textbf{S}patially and \textbf{T}emporally \textbf{R}esolved \textbf{A}dvection as a benchmark.\\
Our new approach will be considered and explained in the following section. The foundation is again the advection of the flow, but in contrast to \textit{PST} it is not relying on an interpolation. For this, however, the approach can only be applied to individual points of the field. In the following, it will be referred to as \textit{ASTRA}.

\section{\textbf{A}pproach using \textbf{S}patially and \textbf{T}emporally \textbf{R}esolved \textbf{A}dvection (ASTRA)}
\label{sec:HybApp}
\begin{figure}[htb]
	\includegraphics[width=0.45\textwidth]{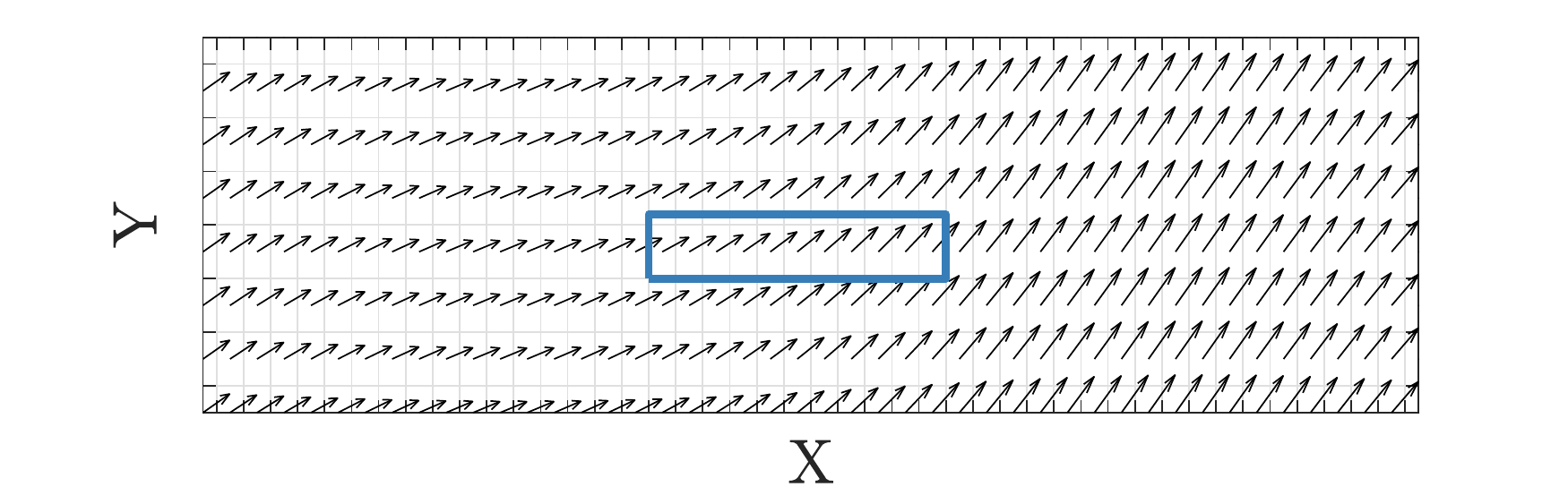}
	\caption{Schematic velocity field of a flow modulation in y-direction. The blue box indicates the extracted vector $\protect\overrightarrow{v}$ from the velocity field at a given position $(x,y)$.}
	\label{fig:VectorfieldTheo}       
\end{figure}

\begin{figure}[htb]
	\includegraphics[width=0.45\textwidth]{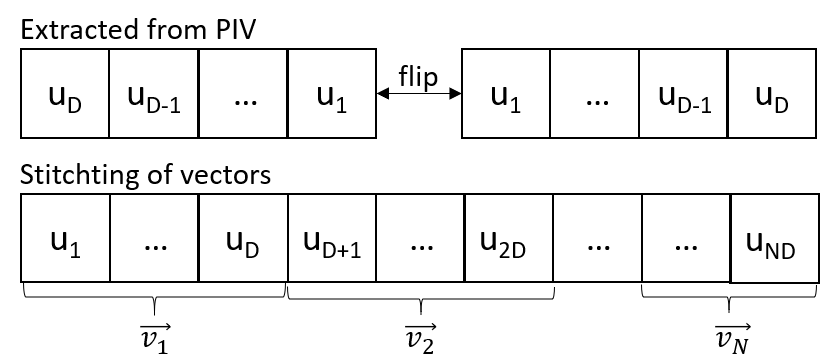}
	\caption{Schematic representation of the composition of the new time series.}
	\label{fig:ExtractionTheo}       
\end{figure}

\begin{figure}[htb]
	\includegraphics[width=0.45\textwidth]{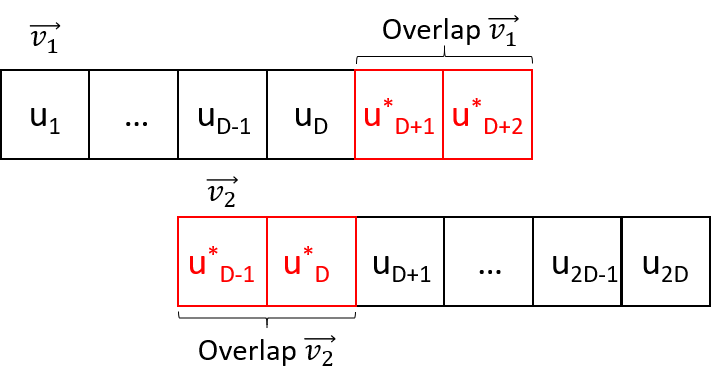}
	\caption{Schematic representation of the overlap during time series stitching. The red boxes correspond to the additionally added entries that provide the overlap.}
	\label{fig:OverlapTheo}       
\end{figure}

Next, we will discuss in more detail the method of \textit{ASTRA}, as a main part of our paper. The influence of different parameters on the resulting reconstructed time series will be investigated. This refers to parameters used during PIV measurement and evaluation as well as the parameters used to generate the time series itself. After optimal parameters are clarified, \textit{ASTRA} is compared to the raw data (\textit{PIV}), as well as \textit{PST}.\\
The general assumption of \textit{ASTRA} is a transport of structures in the flow field with a local mean flow velocity larger than zero, which can be assumed to be stable (or frozen) for a short period of time. Further velocity contributions are assumed to fluctuate around zero. This is one of the first important differences compared to \textit{PST}, which uses the current local velocity at each time step for advection. \textit{ASTRA} uses the local mean flow velocity instead.\\ 
To calculate the displacement $D(x,y)$ of the structures between two successive images at any coordinate $(x,y)$ in the FOV, we use 
\begin{equation}
	D(x,y) = \frac{\overline{u(x,y)}}{F_s\cdot R}.
	\label{eqn:1}
\end{equation}
$D(x,y)$ is hereby given in units of interrogation windows. $\overline{u(x,y)}$ corresponds to the velocity averaged over the measurement period at location $(x,y)$. $F_s$ is the sampling frequency and $R$ the spatial resolution of the PIV measurement.
To form a time series a sequence of the spatial velocity measurements are used to define a vector $\overrightarrow{v} = (\overrightarrow{u_1},...,\overrightarrow{u_D})$ along the main flow direction from each PIV field like it is shown in Fig. \ref{fig:VectorfieldTheo}. Since the displacement $D(x,y)$ can also be a non-integer number, the length of $\overrightarrow{v}$ is rounded. The remainder is added to the subsequent vector length. Assuming an advection of structures from left to right, the last entry of $\overrightarrow{v}$ would be the first sample for the time series. Therefore, $\overrightarrow{v}$ must be flipped to add it to the time series. A sketch of the procedure is shown in Fig. \ref{fig:ExtractionTheo}.
 \\
The new sampled time series has a length of $N \cdot D(x,y)$, where $N$ denotes the number of snapshots taken during measurement. Thus, the sampling frequency increases to
\begin{equation}
	F_{s,HA}(x,y) = \frac{\overline{u(x,y)}}{R}.
\end{equation}
The highest resolvable frequency is therefore limited by the size of the interrogation window, or respectively, the spatial resolution of the PIV $R$ and the mean flow velocity.\\
In time resolved measurements structures of the flow can be found in successive images. The information therefore duplicates and can be used to improve the transition of the merged vectors $\overrightarrow{v}$. Fig. \ref{fig:OverlapTheo} shows such possible overlaps between two extracted vectors. The first-row shows $\overrightarrow{v_1}$. This vector is for example extended by two further entries from the velocity field. These entries, $u_{D+1}^*$ and $u_{D+2}^*$, correspond approximately to the velocities $u_{D+1}$ and $u_{D+2}$ from vector $\overrightarrow{v_2}$. $\overrightarrow{v_2}$ is also extended by two entries $u_{D-1}^*$ and $u_{D}^*$, which correspond to the velocities $u_{D-1}$ and $u_{D}$ from vector $\overrightarrow{v_1}$. Thus, a certain overlap between the individual vectors can be defined.  Hereby, for example, instead of $u_D$ an average value or other weighted sums of $u_D$ and $u_D^*$ can be calculated. The influence of such overlap will be discussed later.
 \\
Since the time series using \textit{ASTRA} are extracted from the velocity fields and their resolution depends on the spatial and temporal resolution of the PIV evaluation, the influence of these experimental parameters will be investigated in the following. The performed hot-wire measurements are used as a comparison.\\ 
\subsection*{Influence of PIV Evaluation Parameter}
\label{sec:PIV_Param}
\begin{figure}[h]
	\includegraphics[width=0.45\textwidth]{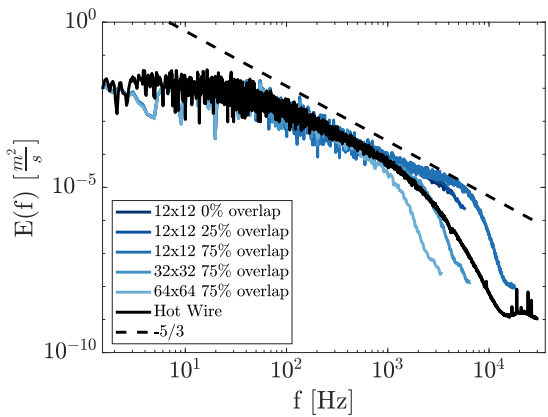}
	\caption{Energy density spectra of time series extracted at $x = 1000$ mm and evaluated using different interrogation window sizes and overlaps during PIV evaluation. The spectra are based on the absolute flow velocity.}
	\label{fig:PIVParam}       
\end{figure}

Having explained \textit{ASTRA}, the influence of the PIV evaluation parameters is examined below. Therefore, the used interrogation window size and overlaps are changed during the PIV evaluation. These parameters determine the spatial resolution of the resulting velocity vector field. Note, the spatial resolution directly determines the new sampling rate. To reduce the influence of coherent structures in this study, we choose a region of homogeneous turbulence at x = $1000$mm. To compare the results, the energy density spectra $E(f)$ of the \textit{ASTRA} time series and the hot-wire measurements are considered.\\
Fig. \ref{fig:PIVParam} shows the resulting energy density spectra. A considerable influence of chosen parameter on the spectra can be observed, which can be attributed to the spatial resolution. IW sizes of 12 x 12 px result in a higher cutoff frequency. Nevertheless, an increase of energy at high frequencies can also be observed due to the evaluation noise. The IW size of 32 x 32 px with an overlap of 75$\%$ show good agreement with the hot-wire. Towards higher frequencies, where also the hot-wire differs from the typical 5/3 spectrum, a low pass filtering effect can be observed. For larger interrogation windows, like 64 x 64 px, high frequencies are underrepresented due to the coarser spatial resolution.\\
Since the results for IW size 32 x 32 px with 75\% overlap best reproduce the hot-wire measurements, these parameters are used for further evaluation.

\subsection*{Influence of Temporal Resolution}
\begin{figure}[h]
	\includegraphics[width=0.45\textwidth]{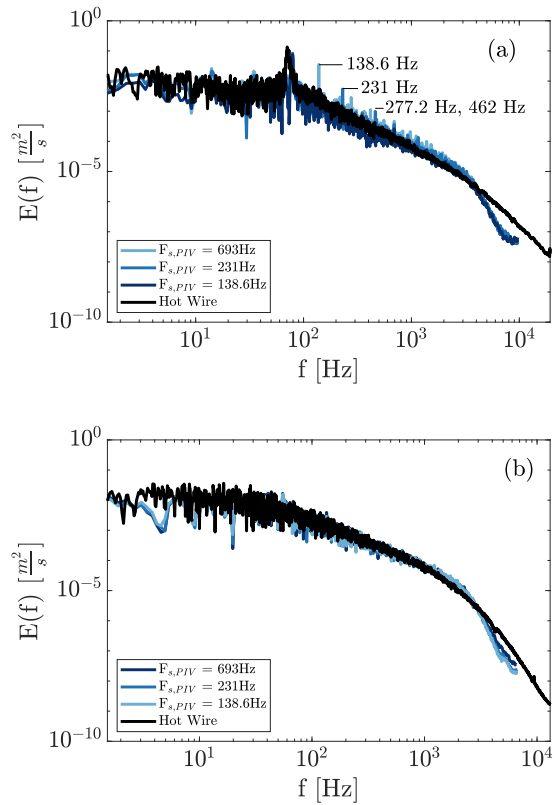}
	\caption{Energy density spectra of time series extracted at (a) $x = 200$ mm and (b) $x = 1000$ mm. The spectra are calculated based on different sampling frequencies $F_{S,PIV}$. The spectra are based on the absolute flow velocity.}
	\label{fig:TempRes_20_100}       
\end{figure}

Another important parameter when performing experiments is the temporal resolution of the PIV measurement. For this reason, the effect of a lower temporal resolution will be investigated. The recorded data set is artificially sampled down by using just every [1,3,5]th velocity field. By reducing the sampling frequency, the displacement of the structures $D$ (see Eqn. \ref{eqn:1}) increases. The length of the extracted vector $\overrightarrow{v}$ will therefore also increase, taking a larger area of the FOV into account.\\
Fig. \ref{fig:TempRes_20_100} shows the resulting energy density spectra at different positions of the flow field for three different sampling frequencies. Fig. \ref{fig:TempRes_20_100}(a) shows the spectra for $x = 200$mm, very close to the grid. 
Due to the meandering jet of the fractal grid, a very dynamic flow situation prevails here, which changes spatially very strongly. The meandering frequency depends on the grid geometry and speed and in the case of this setup is 70Hz-80Hz. In general, all spectra shown in Fig. \ref{fig:TempRes_20_100}(a) agree with the hot-wire. For the lower temporal resolution peaks are showing up at the sampling frequencies and multiples thereof.  This effect is caused by the stitching of the vectors $\overrightarrow{v}$. Due to the changing advection velocity along the increasing vector length, jumps occur at the junction. To avoid such peaks, we therefore recommend that the average velocity $\overline{u(x,y)}$ along $\overrightarrow{v}$ should not deviate more than 5$\%$.\\
In Fig. \ref{fig:TempRes_20_100}(b) the spectra for the homogeneous flow field at $x = 1000$mm are shown. Here the length of the extracted vector $\overrightarrow{v}$ seems to have no influence. Thus, for such flows, it is sufficient if $D$ is smaller than the FOV to perform meaningful spectral analyses.\\

\subsection*{Influence of Vector Overlap}
\begin{figure}[h]
	\includegraphics[width=0.45\textwidth]{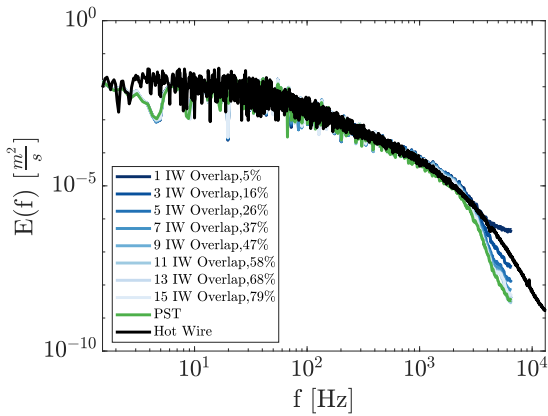}
	\caption{Energy density spectra of time series extracted at $x = 1000$ mm and evaluated using different vector overlaps. The spectra are based on the absolute flow velocity.}
	\label{fig:Overlap}       
\end{figure}

Last, the influence of the only parameter which can be changed in \textit{ASTRA}, namely the overlap between the extracted vectors $\overrightarrow{v}$, will be examined. Fig. \ref{fig:Overlap} shows the energy density spectra for different vector overlaps. The shade of blue indicates the overlap size. Vectors $\overrightarrow{v}$ extracted at the position $x=1000$mm have an average movement of 15.1mm or $D(x,y)=19$ IWs between two snapshots. The overlap is therefore also given as a percentage of the total advection $D(x,y)$in between frames.
 \\
The overlap shows only an influence for high frequencies above $4$kHz. Here the effect of a low pass filter for overlaps $>10\%$ can be observed. This justifies the use of such overlaps to smooth the transition between vectors. Since a small overlap leads to additive noise in high frequencies and too large an overlap low pass filters the data, an overlap of three IWs is used in the following. \\
For a direct comparison, the figure also contains \textit{PST}. The \textit{ASTRA} spectra can be seen to converge with increasing overlap from the hot-wire spectrum towards \textit{PST}. This indicates a strong low pass filtering effect of \textit{PST}. \\ 
In summary, \textit{ASTRA} shows good results for very inhomogeneous flows (close to the grid), as well as for very homogeneous ones. By artificially reducing the sampling frequency a limit for the temporal resolution was found for strongly inhomogeneous flows. If the temporal resolution is not sufficiently high artifacts will occur in the spectrum. For homogeneous flow fields, this restriction seems to be less severe. It could be demonstrated that the use of an overlap has a positive effect on the merging of the time series. A too large overlap acts as a low pass filter and the results converge towards \textit{PST} by Scarano and Moore. In this way, the dynamics of the small scales can be preserved by keeping the overlap as small as possible.

\section{Comparison of Results} 
\label{sec:Results}
Having introduced the approaches for generating PIV spectra and time series and described our proposed \textit{ASTRA} method, these are now applied to experimental data behind the fractal grid in the following section. Reference should be made to the supplementary material containing the time-independent statistical analysis of the measured data. Mean value, standard deviation, skewness, and kurtosis of the velocity measurements along the center line can be found there. It is shown that the PIV and hot-wire provide comparable results. \\
In this part, we first compare the plain time series extracted from the PIV measurement itself, using \textit{ASTRA} and \textit{PST}. 
Based on this time series the energy density spectra are calculated. Finally, the increments of the time series are analyzed. These directly reflect the structure functions of the present flow.
\subsection*{Generated Time Series}
\begin{figure}[h]
	\begin{centering}
		\includegraphics[width=0.45\textwidth]{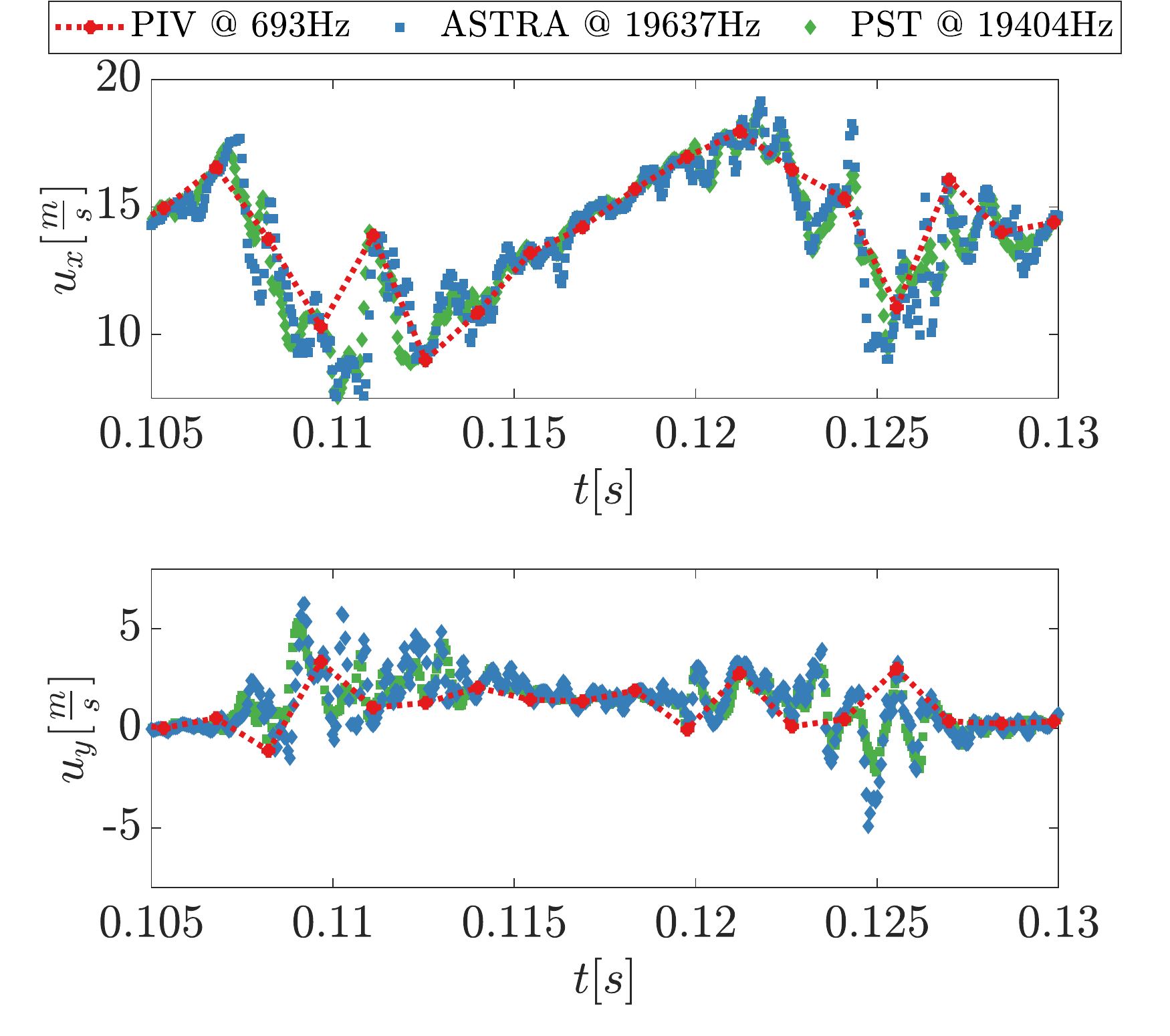}
		\caption{Time series extracted from the PIV measurement at $x=200mm$. The red dots represent the data from the PIV measurement itself. The dashed lines show the results for both advection-based approaches. Fig. (a) shows the $u_x$ component of the flow velocity and Fig. (b) the $y$ component.}
		\label{fig:Timeseries}       
	\end{centering}
\end{figure}

Before a statistical analysis is carried out, first the generated time series should be compared. In Fig. \ref{fig:Timeseries} the $u_x$ and $u_y$ components of the flow velocity at the centerline are shown for the position $x=200$mm in the very dynamic area behind the grid. The red dots represent the data points obtained using the raw \textit{PIV} data with given frame rate. The blue dots represent the new generated time samples using \textit{ASTRA} and the green ones the time samples generated using \textit{PST}.\\ 
Both the \textit{PST} approach and \textit{ASTRA} proposed here capture the same trend as the underlying \textit{PIV} data. Both approaches create time series with a sampling rate of nearly $20$kHz, almost thirty times higher than the initial data. At first sight, the advection-based approaches show very similar behavior. However, a closer look reveals \textit{ASTRA} to show higher dynamics. The approach shows significantly higher maxima and minima compared to \textit{PST}. This applies to both velocity components.\\
To examine whether this is the case, and which influence this difference has on the statistics of the time series, spectra and an increment analysis will be used in the following.

\subsection*{Spectra and Increments}
\begin{figure}[h]
	\includegraphics[width=0.45\textwidth]{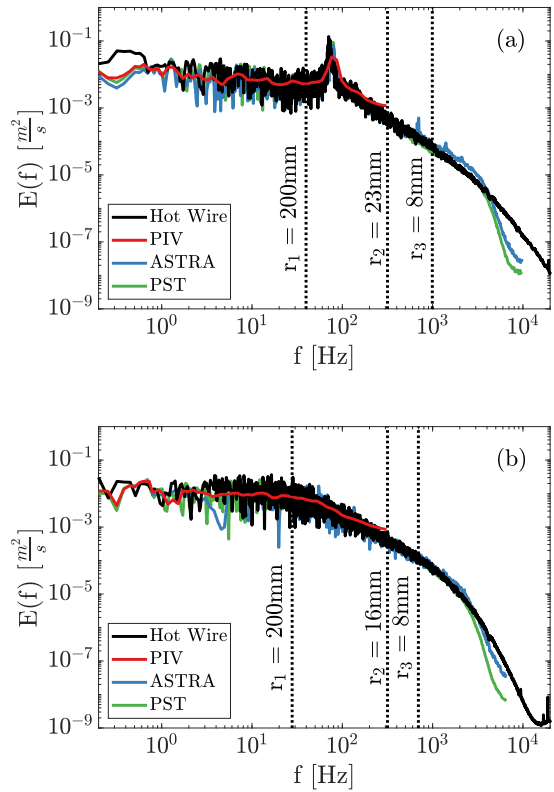}
	\caption{Energy density spectra calculated from the different time series approaches at (a) $x=200$mm and (b) $x=1000$mm. Also shown are the spatial scales $r$ chosen for the increment analysis.}
	\label{fig:Spectra}       
\end{figure}

Following, the focus is put on spectra and increment analysis. Such higher order statistics strongly depend on the temporal resolution of the measurement data. The increments of a time series for example directly represent all structure functions describing the flow field. Therefore, if they differ, the actual statistics are not reflected sufficiently leading to a misrepresentation of the given flow.\\
First, the spectra calculated from the previous shown time series will be compared for the two investigated distances. A very significant feature of the fractal grid is the meandering jet in the near grid region. At the present wind speed, the jet meanders at about $70$Hz up to $80$Hz. Due to the mixing of the flow, the intensity of the jet frequency decreases with increasing distance from the grid. For this reason, the spectra at the position $x=200$mm behind the grid show the characteristic peak (see Fig. \ref{fig:Spectra}(a)). For the position further downstream, this frequency is no longer found (see Fig. \ref{fig:Spectra}(b)). \\
Comparing the spectra of the different approaches to extract a time series in Fig. \ref{fig:Spectra} show a similar shape for the dynamic case close to the grid (Fig. \ref{fig:Spectra}(a)). There is not much difference between our new \textit{ASTRA} approach and the \textit{PST} approach. Just a slightly stronger low pass filtering effect seems to be present for the \textit{PST} approach. Also, the spectral gain compared to the raw \textit{PIV} data can be seen.\\
For the more homogeneous turbulent flow at a distance of $x=1000$mm, a similar picture emerges (see Fig. \ref{fig:Spectra}(b)). Here, however, a significantly stronger low-pass filter effect can be observed for the \textit{PST} approach. \textit{ASTRA}, on the other hand, follows the spectrum of the hot wire up to very high frequencies.\\
This spectral analysis again underlines the advantage of the advection-based methods compared to using the raw data itself. They allow to obtain much more detailed spectra from the raw data.\\

\begin{figure*}[htb]
	\begin{centering}
		\includegraphics[width=0.95\textwidth]{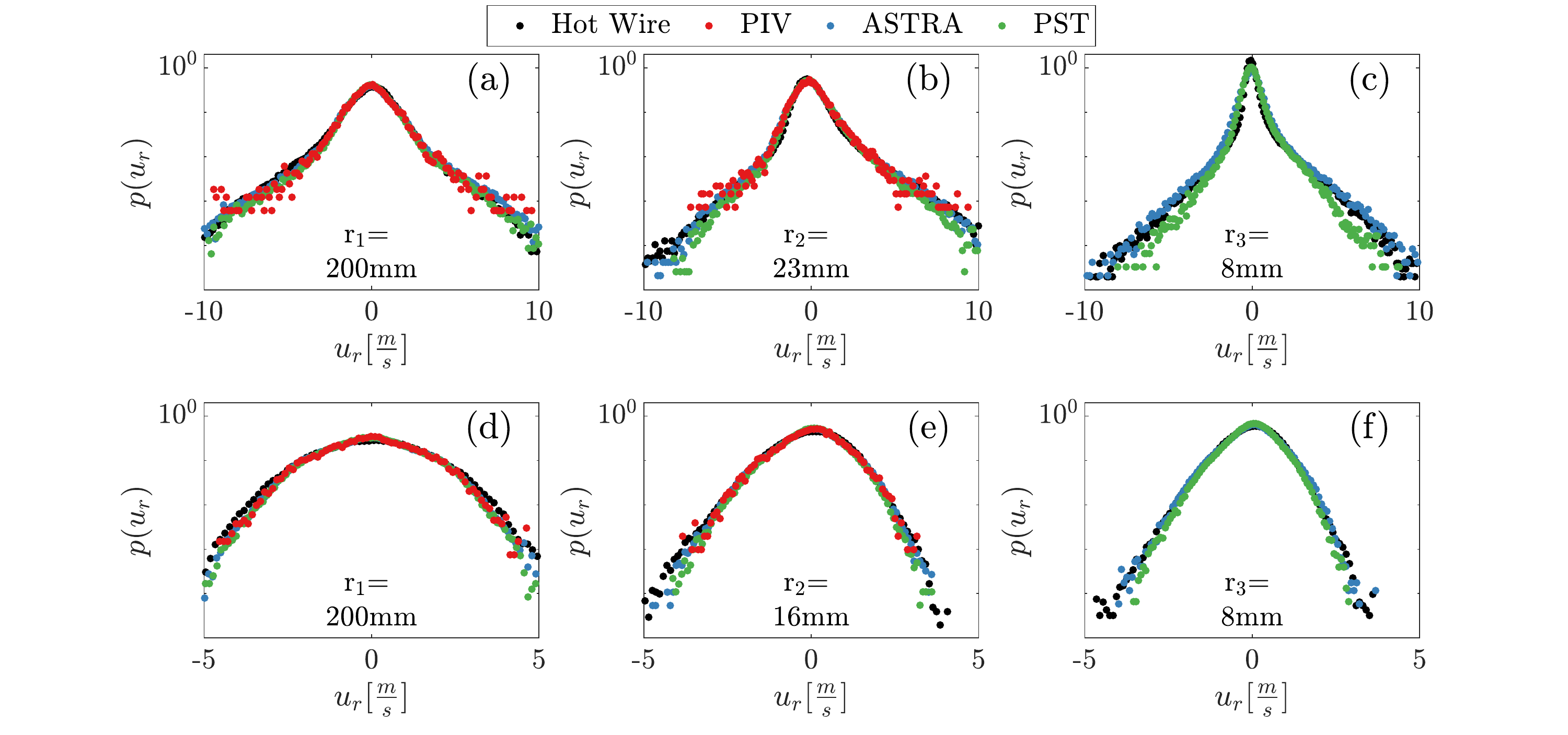}
		\caption{Probability of the increments $u_r$ for different length scales $r$. (a)-(c) Represent the increments calculated from data at $x=200$mm. (d)-(f) Represent the increments calculated from data at $x=1000$mm. The increments are chosen corresponding to the FOV size ((a),(d)), the given PIV sampling frequency of 693Hz ((b),(e)) and a scale slightly above the dissipation length of the flow ((c),(f)).}
		\label{fig:Increments20}       
	\end{centering}
\end{figure*}

To further deepen the insight into the statistics of the flow, the increments shall also be considered to check whether the two-point statistics are sufficiently well captured by the various approaches.\\ 
An increment is defined as difference between two measured velocities at time $t$ and $t+\tau$ or in space the position $x$ and $x+r$. A temporal or spatial increment is defined as follows:
\begin{equation}
u_\tau(t) = u(t) - u(t+\tau),
\end{equation}
\begin{equation}
u_r(x) = u(x) - u(x+r).
\end{equation}
The probability distribution $p(u_r)$ of the increments is shown for different length scales in Fig. \ref{fig:Increments20} for a position close to the grid (a)-(c) and for a position further downstream (d)-(f). The length scales are chosen to represent the size of the FOV (left), the sampling frequency of the PIV measurements (middle) and a scale slightly above the dissipation length of given flow (right). The selected scales can also be found as dashed lines in the spectra (See Fig. \ref{fig:Spectra}).\\
Comparison of increments in the dynamic flow at $x=200$mm show agreements between hot-wire, \textit{PIV}, \textit{ASTRA} and \textit{PST} for large scales (see Fig. \ref{fig:Increments20}(a)).  For smaller scales in the range of the sampling frequency of the \textit{PIV} measurement (Fig. \ref{fig:Increments20}(b)), however, the \textit{PST} approach shows a decrease in the tails of the distribution compared to the other methods. For the smallest scale shown (Fig. \ref{fig:Increments20}(c)), this deviation is even more severe. For \textit{ASTRA} such deviation is not observed. The increments agree very well with those of the hot-wire measurement for all scales. For the smallest scale, increments can no longer be determined from the raw \textit{PIV} data, as this is below temporal resolution.\\
Similar observations can also be made for the increments from homogeneous turbulence(Fig. \ref{fig:Increments20}(d-f)). Here, for large scales, a good agreement is also shown. At smaller scales, the \textit{PST} approach deviates from the hot-wire measurements, while \textit{ASTRA} provides the same statistics as the hot-wire. At this position, however, the deviations are significantly smaller.
 \\
The incremental two-point statistics indicates \textit{ASTRA} to provide reliable results for all scales in the flow. The \textit{PST} approach, on the other hand, deviates in two-point statistics for scales belonging to frequencies above the sampling frequency, despite matching spectra. This indicates the interpolation to distort the properties of the multi point statistics in this approach. Since the increments directly reflect the structure functions, it is evident from this analysis that the \textit{PST} approach is not suitable for deeper analyses of the turbulent structures, while \textit{ASTRA} presented here does.

\section{Conclusion}
\label{sec:Conclusion}
In this study, a new advection-based approach is presented to extract high temporal resolution time series from PIV measurements by combining temporal and spatial information contained in PIV measurements. The approach is compared to an already known advection-based approach by Scarano and Moore \cite{Scarano2012}. \textit{ASTRA} is validated using a complex flow. Hereby, the reliability of the approach is verified in different flow situations. Compared to the \textit{PST} approach, \textit{ASTRA} provides only local time series for a selected position in the PIV field and no velocity fields. However, the process can be repeated for any point, making all information accessible. Another advantage of \textit{ASTRA} is the use of only actual measurement data, which makes artificial interpolation of the data unnecessary. This saves memory and time during the extraction of the time series, since the interpolation of the new velocity fields used in \textit{PST} involves a large computational effort.\\After the procedure for generating the time series is presented, the optimal parameters using \textit{ASTRA} are defined based on spectral analyses. Particularly important are the interrogation window properties selected during the PIV evaluation. The chosen parameter directly influences the spatial resolution and can add noise to the results if chosen incorrectly. In addition to parameters given by the measurement, the influence of an overlap used when compiling the time series is investigated. If this overlap used in \textit{ASTRA} is increased, the resulting time series converges to the Scarano and Moore approach for an overlap of 100$\%$.\\
In further statistical analysis \textit{ASTRA} also provides plausible results. This holds true for extracted spectra and increment analysis. The incremental analysis proved \textit{ASTRA} to capture velocity fluctuations on small scales better compared to the \textit{PST} approach. This is a decisive criterion, especially about further analyses, such as entropy analysis \cite{Fuchs2020}, since it is precisely the fluctuations on small time scales that must be captured sufficiently well.\\
For these reasons, \textit{ASTRA} shown here represents a very useful extension of statistical analysis of turbulence with PIV. The method allows a highly time resolved analysis and the extraction of higher order statistics from the flow even with limited temporal resolutions.
\bibliographystyle{spphys}       
\bibliography{D:/Clouds/ownCloud_Arbeit/Paper_Drafts/High_Resolved_Spectra_From_PIV/LaTeX_DL_468198_240419/PIV_Spectra}   

\clearpage


\section*{Supplementary material}

**Supplementary Material**

In the supplementary material, the non-time-dependent statistics of the flow behind the fractal grid are presented. These results are intended to emphasize the correctness of the measured flow field. It will also be shown that the PIV and hot-wire measurements show the same basic statistics, justifying the comparison between these methods in the paper.

For simplicity, the time series were determined using *ASTRA* and *PST* approach only at the locations from the *PIV* data where hot-wire measurements were also performed.

One Point Statistics Along Centerline

In Fig. 1(a) the profile of the absolute velocity along the center line is shown. The PIV measurement shows the same decay of the velocity as the hot wire. This decrease behind the fractal grid is a typical feature. Fig. 1(b) shows the difference between the *PIV* measurement itself and the statistics calculated based on the different approaches. Although the data basis is the same and the approaches should provide the same results as the *PIV* measurement itself, slight deviations can be seen. It shows that both underestimate the mean velocity, but *ASTRA* is closer to the results of the *PIV* measurement.

Fig. 2(a) shows the standard deviation along the centerline. Again, PIV and hot-wire are in good agreement. The standard deviation increases until the maximum is reached at $x = 350$mm, as expected. The PIV results again fit well with the hot-wire measurements. Fig. 2(b) again shows the difference between the PIV data set and the approaches. As with the mean value, deviations between the underlying data and the extracted time series can be seen here. Especially in the highly dynamic range close to the grid, *ASTRA* overestimates the standard deviation. In the more homogeneous areas, the values agree well. For the *PST* approach, a systematic underestimation of the standard deviation is found for the entire range.

The third moment, the skewness, is plotted in Fig. 3(a). Again, the PIV and hot-wire measurements give the same results. The skewness shows the minimum typical for the fractal grid. Fig. 3(b) again shows the difference between *PIV* and the used approaches. Both approaches show deviations in the near range of the grid. The deviations are approximately the same for both approaches.

Finally, the kurtosis with its typical sharp peak is shown in Fig. (a) for PIV and hot-wire. The differences between PIV measurement and extracted time series is shown in Fig. 4(b). Both approaches show similar deviations.

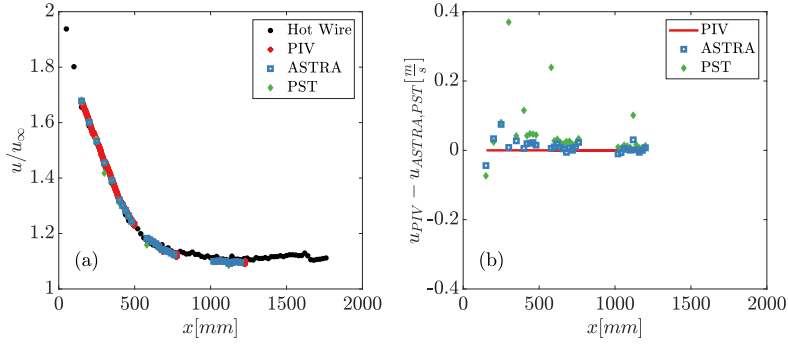

**Fig. 1** (a) Mean absolute velocity along the center line behind the fractal grid for the different approaches. (b) Difference between raw *PIV* measurement and the mean value calculated from the time series given by the different approaches.

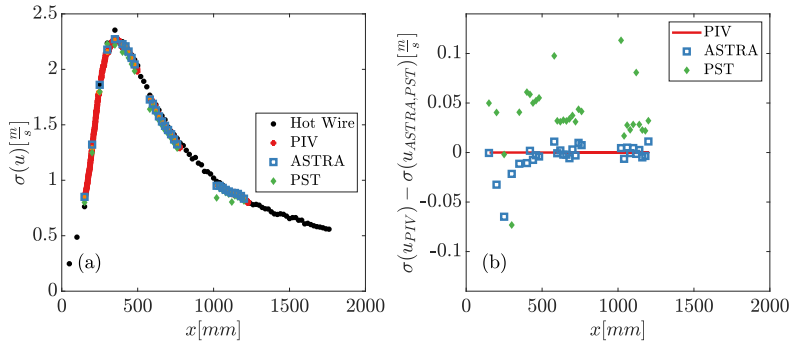

**Fig. 2** (a) Standard deviation of the measured absolute velocity along the centerline behind the fractal grid. (b) Difference between raw *PIV* measurement and the standard deviation calculated from the time series given by the different approaches.

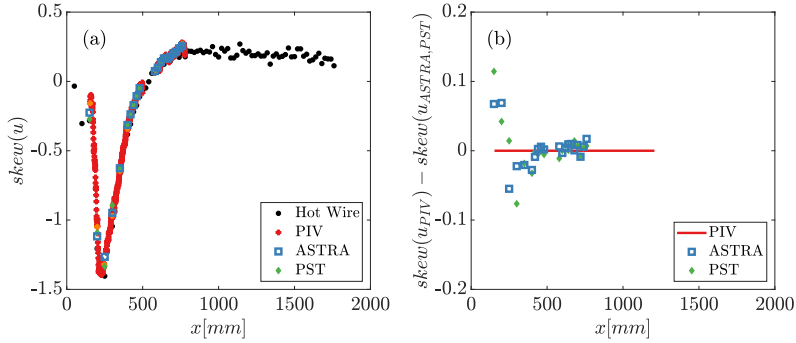

**Fig. 3** (a) Skewness of the measured absolute velocity along the centerline behind the fractal grid. (b) Difference between raw *PIV* measurement and the skewness calculated from the time series given by the different approaches.

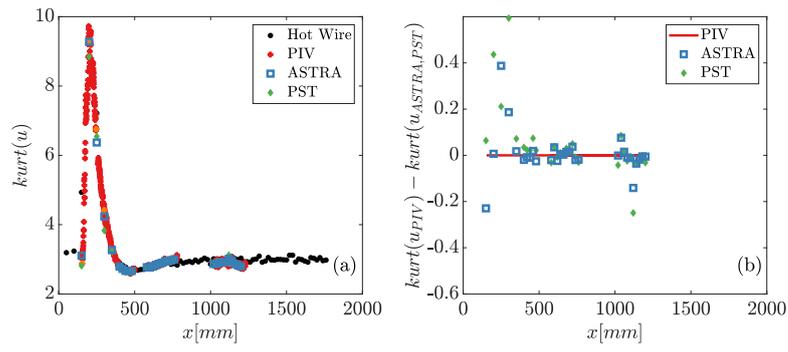

**Fig. 4** (a) Kurtosis of the measured velocity along the centerline behind the fractal grid. (b) Difference between raw *PIV* measurement and the kurtosis calculated from the time series given by the different approaches.



\end{document}